\documentstyle[12pt,epsfig]{article}                
\def\lapp{{\ \lower 0.6ex \hbox{$\buildrel<\over\sim$}\ }}
\def\gapp{{\ \lower 0.6ex \hbox{$\buildrel>\over\sim$}\ }}
\begin{document}
\begin{titlepage}
\vspace*{-1cm}
\begin{flushright}
FTUV/98/27 \\
IFIC/98/27  \\
March 1998 \\
\end{flushright}                                
\vskip 1.cm  
\begin{center}                                                                  
{\Large\bf Fermiophobic and other non--minimal neutral Higgs bosons at the LHC}\\
 \vskip 1.cm
{\large A.G.~Akeroyd\footnote{akeroyd@flamenco.ific.uv.es}}\\
Dept. de Fisica Teorica, IFIC/CSIC\\
Univ. de Valencia, 46100 Burjassot, Valencia, Spain \\
\vskip 1cm                                                                    
\end{center}

\begin{abstract}
The phenomenology of neutral Higgs bosons from non--SUSY,
extended Higgs sectors is studied in the context of the LHC, 
with particular attention given to the case of a fermiophobic Higgs.
It is found that enhanced branching ratios to $\gamma\gamma$ and
$\tau^+\tau^-$ are possible and can provide clear signatures, while detection
of a fermiophobic Higgs will be problematic beyond a mass of 130 GeV.  
\end{abstract}
\hspace*{1.0cm}

\vfill
\end{titlepage}                                                                 
\newpage
\section{Introduction}
It is well known that the Standard Model (SM) \cite{Wein} requires the breaking
of the $SU(2)\times U(1)$ symmetry. Introducing a complex scalar doublet
with a non--zero vacuum expectation value (VEV) is an elegant way of achieving
this, and predicts one neutral scalar -- the Higgs boson ($\phi^0$)
 \cite{Higgs}. Models 
with $N$ doublets, which we shall call `multi--Higgs--doublet models' (MHDM) 
are possible \cite{Gross}, 
and in particular two doublets are required for the minimal Supersymmetric 
Standard Model (MSSM) \cite{Gun}. One of the main goals at future
colliders will be to 
search for  
the above Higgs bosons, with previous null searches at LEP having produced the 
bounds
of $M_{\phi^0}\ge 70.7$ GeV \cite{Sop} for the minimal SM, and $M_{h^0}\ge 65.2$ 
GeV \cite{Del} for the
lightest CP--even neutral Higgs boson ($h^0$) of the MSSM. 

Although there are
many theoretically sound extended Higgs models, most attention is given to
the minimal SM and the MSSM. We shall
be considering a non--SUSY, non--minimal SM in which only the Higgs sector
is enlarged and any new physics is assumed to enter at a higher
energy scale. Thus the low energy structure resembles the SM with
an extended Higgs sector. In an earlier publication \cite{Ake3} we considered
the phenomenology of these models at LEP2, assuming that the lightest 
CP--even neutral scalar ($h_1$) was the only Higgs boson in range at this 
collider. It was shown that detection is possible and parameter spaces
exist for distinctive signatures. We recall that distinguishing
among the many possible Higgs representations is a key issue at
future colliders, and for the purpose of this paper we shall assume that
$\phi^0$ and $h^0$ possess a very similar phenomenology - distinguishing
between these two particles provides a challenge for future colliders and
is discussed elsewhere \cite{Hab}. In this paper we 
study the phenomenology of $h_1$ at the LHC in order to find out whether 
discovery/distinctive signatures are again possible, assuming that $h_1$ 
escaped detection
at LEP2 and therefore its mass is constrained, $M_{h_1}\ge 100$~GeV.
A caveat here is that this bound is for $h_1$ with couplings to
vector bosons of $\phi^0$ strength, but if the
vector boson ($h_1VV$) coupling is very suppressed a relatively light 
$h_1$ may escape detection at LEP1 and LEP2 \cite{Mar}, \cite{Kraw}.
 
Throughout the paper we highlight the existence of a fermiophobic
Higgs ($H_F$) \cite{Hab1}$\to$ \cite{Ake4}  which is possible in the models
 we consider. Such 
a particle has been searched for recently at both the Tevatron \cite{Fermtev}
 and
LEP \cite{FermLep}, and is currently bounded to be heavier than 81 GeV (95$\%$).
 We note
that these bounds are for an $H_F$ with $\phi^0$ strength coupling to
vector bosons, although in general this is not the case. Our
work is organized as follows. In Section 2 we introduce the relevant 
non--minimal Higgs models and investigate their couplings to the fermions
and gauge bosons. Section 3 examines 
their phenomenology at the LHC in four separate decay channels, and finds
parameter spaces which exhibit signals not possible for $\phi^0$ or $h^0$. 
 Finally, Section 4 contains our conclusions.
         
\section{Extended Higgs Sectors}

The minimal SM consists of one Higgs doublet ($T=1/2$, $Y=1$), although 
extended models can be considered and have received substantial 
attention in the literature. For a general review see Ref. \cite{Gun}. 

The theoretical structure of the two--Higgs--doublet model (2HDM) is well 
known, while the general multi--Higgs--doublet model (MHDM) 
 \cite{Gross}, \cite{Ake} 
has received substantially less attention, and usually only
in the context of the charged Higgs sector.
In this paper we shall be considering the models 
of our earlier work i.e. we shall again focus on the four distinct
versions of the 2HDM (denoted Models I,I$'$,II and II$'$)
 with mention given to the MHDM when appropriate. 
The Higgs sector of the MSSM requires Model~II type 
couplings and thus the phenomenology of Model~II has received the most
attention in the literature. Models I$'$ and II$'$ are rarely 
mentioned, while Model I has received limited attention.
 We shall be considering the
lightest CP--even Higgs scalar ($h_1$) of the above models, and for the 2HDM 
its couplings to the fermions are given in Table~1.

\begin{table}[htb]
\centering
\begin{tabular} {|c|c|c|c|c|} \hline
& Model~I & Model~I$'$ & Model~II & Model~II$'$  \\ \hline
$hu\overline u$ & $\cos\alpha/\sin\beta$ & $\cos\alpha/\sin\beta$
& $\cos\alpha/\sin\beta$  & $\cos\alpha/\sin\beta$      \\ \hline
$hd\overline d$ & $\cos\alpha/\sin\beta$& $\cos\alpha/\sin\beta$ 
&$-\sin\alpha/\cos\beta$ &$-\sin\alpha/\cos\beta$      \\ \hline
$he\overline e$ &  $\cos\alpha/\sin\beta$  & $-\sin\alpha/\cos\beta$ 
  &$-\sin\alpha/\cos\beta$ 
&  $\cos\alpha/\sin\beta$\\ \hline
\end{tabular}
\caption{The fermion couplings of $h_1$ in the 2HDM relative to those for the
 minimal SM Higgs boson ($\phi^0$).}
\end{table}
\noindent
Here $\alpha$ is a mixing angle in the neutral Higgs sector
and $\beta$ is defined by $\tan\beta=v_2/v_1$ ($v_i$ is the VEV of the $i^{th}
$ doublet and $v^2=\sum_{i=1}^N v_i^2=(246$ GeV$)^2$). In the MSSM, which 
is a constrained
version of the 2HDM (Model~II), the angles $\alpha$ and $\beta$ are
correlated. For the models that we shall consider $\alpha$ and $\beta$ are 
independent. In all the models there exists a bound on $\tan\beta$ 
from considering the effects of the charged Higgs on the $Zb\overline b$
vertex \cite{Hollik}. Although the full $H^{\pm}tb$ coupling  depends on the model
there is a piece $m_t\cot\beta$ which is common to all models and
dominates unless $\tan\beta$ is very large. From this vertex $\tan\beta$ 
may be constrained from current experimental data to be 
(for $M_H=M_Z$, where $M_H$ is the mass of $H^{\pm}$): 
\begin{equation}
\tan\beta\ge 1.54\;(95\%\;c.l)
\end{equation} 
This improves the bound $\tan\beta\ge 2$ which was found in Ref. \cite{Park},
with
the 1995 value for $R_b$. We have used the graphs in Ref. \cite{Hollik} to
obtain the bound in Eq. (1).  In our previous work we used $\tan\beta\ge 1.25$.
For heavier $M_H$ smaller values of $\tan\beta$ are allowed.
Note that we have used $M_H=M_Z$ which is permissible in Model I and I$'$, 
and in the MHDM \cite{Ake}. In Model II and II$'$ $M_H$ is constrained by 
$b\to s\gamma$
to be greater than $244+63/(\tan\beta)^{1.3}$ GeV \cite{CLEO}. 
Throughout the paper we shall be varying $\beta$ from $\pi/4\to \pi/2$,
that is, we shall use a lower bound of $\tan\beta=1$.
The angle $\alpha$ may be varied
from $-\pi/2\to \pi/2$, although in our analysis it is sufficient
to vary $\alpha$ from $0\to \pi/2$. 

From Table~1 one can vary the angles $\alpha$ and $\beta$ independently in
order 
to find parameter spaces for extreme branching ratios (BRs). There are
differences here from the LEP2 scenario \cite{Ake3} due to the rapid 
strengthening
of the $VV^*$ channel for $M_{h_1}\ge 100$ GeV, these decays being
negligible at LEP2. In addition the LHC allows different production
mechanisms which will be scaled by mixing angle factors. We note that
the $h_1VV$ coupling in all 4 models is scaled relative to that of 
$\phi^0VV$ by a factor $\sin(\beta-\alpha)$.

\section{Phenomenology at the LHC}
At the LHC the
dominant production process for $\phi^0$ is that of $gg$ fusion, 
which proceeds via a top quark loop \cite{ggH}, \cite{LHC}, \cite{Zer}. Other 
production mechanisms are
considerably smaller, with the next largest being $WW$ fusion. 
Vector boson fusion \cite{WWH} is
always suppressed in a 2HDM by a factor $\sin^2(\beta-\alpha)$, 
while $gg$ fusion
can be enhanced by up to a factor of two. If the $gg$ fusion process
is absent then the cross--section $\sigma(pp\to h_1X)$ is heavily
diminished, and this happens in the case of a fermiophobic Higgs. 
The Higgs bremsstrahlung off a $b$ quark is small for $\phi^0$ but in Model
II and II$'$ it can be boosted for large values of $\tan\beta$ and may become
the dominant production process.            

The most studied channels in the literature for detecting $\phi^0$ are
\cite{LHC}, \cite{0-74}:
\begin{itemize}
\item[{(i)}] $\phi^0\to \gamma\gamma$ for 80 GeV$\le M_{\phi^0}\le 130$ GeV. 

\item[{(ii)}] $\phi^0\to ZZ^{(*)}\to llll$ for 130 GeV$\le M_{\phi^0}\le 800$ 
GeV. 
\end{itemize}
Other channels are considered to give a lesser chance of detection, e.g.
those which make use of $\phi^0\to b\overline b$ decays are swamped
by backgrounds.
For the models that we shall consider there are four possible channels which
could distinguish $h_1$ from $\phi^0$:
 \begin{itemize}
\item[{(i)}] An enhanced signal in channels which exploit the
 decay $h_1\to \gamma\gamma$. 

\item[{(ii)}] An enhanced signal in the channel $h_1\to \tau^+\tau^-$.

\item[{(iii)}] An enhanced signal in the channel $h_1\to t\overline t$ or 
$b\overline b$ in the heavy Higgs mass region.

\item[{(iv)}] An enhanced or suppressed signal in the channel 
$h_1\to ZZ^{(*)}\to llll$. 
\end{itemize}
In the subsections that follow we shall see that some of the above signatures
 are exclusive to a particular 2HDM
model, while others only suggest that a detected neutral Higgs boson
would be of a non--minimal nature.  
Each of the above signatures will be examined in the
context of $pp$ collisions at $\sqrt s=14$ TeV and ${\cal L}=30$
or $100$ fb$^{-1}$. We shall focus on the four distinct versions of the 2HDM,
remembering that $h_1$ from the MHDM can mimic any $h_1$ from the 2HDM.
However, there are possible differences in the phenomenology of the
2HDM and MHDM. To illustrate this, we note here that once 
$\alpha$ and $\beta$ have been chosen to obtain a distinctive BR, 
the production cross--section is constrained and often suppressed compared
to $\phi^0$. This is due to the fact that  if (say) 
large $\cos\alpha\to 1$ is required to enhance a particular partial width, any
coupling proportional to $\sin\alpha$ will be automatically reduced 
(see Table~1).
In the MHDM this correlation among the couplings is relaxed,
 as explained in Ref. \cite{Ake3},
due to the more complicated mixing matrices and the presence of more
VEVs.
Hence it is possible to have the same enhanced BRs in the MHDM along with
a larger production cross--section, resulting in more signal events overall.    

\subsection{$h_1\to \gamma\gamma$}
For $\phi^0$, the decay to $\gamma\gamma$ 
proceeds via charged particle loops (see Fig.~1),
 i.e. fermions and $W$ bosons. 
\begin{figure}[htbp]
\begin{center}
\mbox{\mbox{\epsfig{file=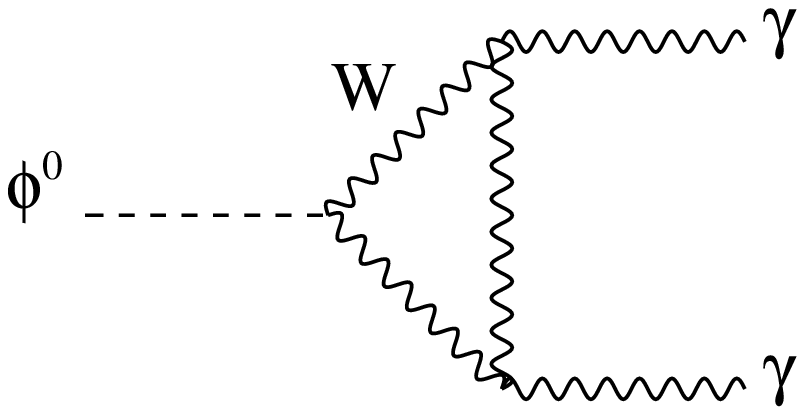,height=2.5cm}}
\mbox{\epsfig{file=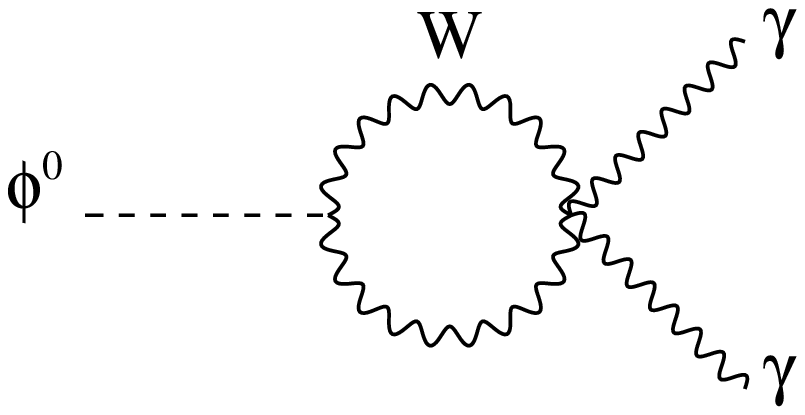,height=2.5cm}}
\mbox{\epsfig{file=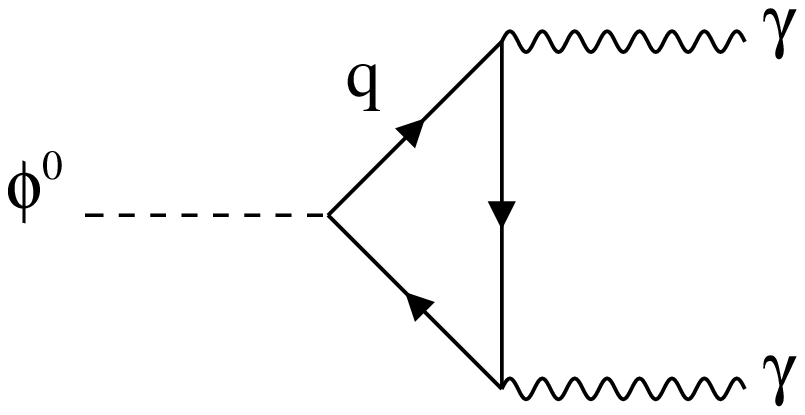,height=2.5cm}}}
\end{center}
\vspace{-5mm}
\caption{Decays of $\phi^0\to \gamma\gamma$.}
\label{Fig:Fig1}
\end{figure} 
We thus concentrate
on the decay $\phi^0\to \gamma\gamma$ and look for ways of enhancing
$h_1\to \gamma\gamma$. For $\phi^0$ the 
vector boson loops dominate and contribute with opposite sign to the
fermion loops. One--loop corrections are small \cite{Zer}, and the tree--level
 width can be written  as \cite{Gun}:
\begin{equation}   
\Gamma(\phi^0\to \gamma\gamma)={G_F\alpha^2M_{\phi^0}^3\over 128\sqrt 2\pi^3}
|\sum_{f} N_cQ^2_fA_f(\tau_f)+A_W(\tau_{W})|^2\,,
\end{equation}
where $N_c$ is the colour factor, $Q_f$ is the electric charge of the fermion,
and the $\tau$ variables are defined by
\begin{equation}
\tau_f={M^2_{\phi^0}\over 4m^2_f}\;\;\;{\rm and}\;\;\;
\tau_W={M^2_{\phi^0}\over 4M^2_W}\;.
\end{equation}

The above equations show that the dominant fermion loop is that of the
$t$ quark (i.e. $f=t$). The amplitudes ($A$)
 are real and vary from $A_W=-7 (-12)$ for $\tau=0(1)$, $A_f=4/3(2)$ for 
 $\tau=0(1)$.
 Hence the $W$ loops are dominant. For the minimal SM Higgs, 
 BR~$(\phi^0\to \gamma\gamma)\approx 0.1\%\to
 0.2\%$ for 80 GeV $\le M_{\phi^0}\le 130$ GeV. To increase 
 this BR for $h_1$ one would naively wish to enhance the width 
 $\Gamma(h_1\to \gamma\gamma)$ significantly, but this is not an option
 since:
\begin{itemize}
\item[{(i)}] The $W$ loop contributes the most, and the $h_1WW$ vertex is
suppressed relative to $\phi^0WW$ by a factor $\sin^2(\beta-\alpha)$.

\item[{(ii)}] Although it is possible to reduce the strength of
the $t\overline t$ vertex and thus 
cause less destructive interference, this does not  
increase the width much, and the corresponding reduction in 
$\sigma (gg\to h_1)$ is disastrous from the point of
view of the total cross--section. 

\end{itemize}
A better way of increasing BR~$(h_1\to \gamma\gamma)$ is to suppress other
decay channels such as $b\overline b$ (which dominates in the relevant mass
region). This suppression can be achieved in Models II
 and II$'$ for
small $\sin\alpha$ although in
Model II the above parameter choice 
 suppresses $b\overline b$ and $\tau^+\tau^-$ decays
simultaneously, thus further enhancing BR~$(h_1\to \gamma\gamma)$.
 In the extreme case of 
BR~$(h_1\to b\overline b)\to 0$ ($\sin\alpha\to 0$), 
one finds that     
enhancements of a factor two (relative to $\phi^0$) are possible for 
BR~$(h_1\to \gamma\gamma$). 
Choices of $\tan\beta$ close to 1 would cause a factor 2 increase
for the the main production mechanism
 $gg\to h_1$ relative to that for $\phi^0$.  Table~2 
(from Ref. \cite{0-48}) shows the expected 
signal for $\phi^0$ for ${\cal L}$=30 fb$^{-1}$
with the significance being rather low in the range 
80 GeV$\le M_{\phi^0}\le 100$ GeV.
For $h_1$ the enhancement of $\gamma\gamma$ events by a factor up to 4 would 
provide a clear signature, and Table~3 shows the
signal numbers for Model II with
$\alpha=0$ and $\beta=\pi/4$.
\begin{table}[htb]
\centering
\begin{tabular} {|c|c|c|c|c|c|c|} \hline
$ M_{\phi^0}$ (GeV) & 80  & 100 & 120   \\ \hline
 BR~($\phi^0\to \gamma\gamma$) &$0.09\%$   & $0.15\%$  & $0.23\%$ \\ \hline 
 $S/\sqrt B$ & 1.5  &2.7 & 4.1 \\ \hline
 \end{tabular}
\caption{Signals for process $pp\to \phi^0X$, $\phi^0\to 
\gamma\gamma$, for ${\cal L}$=30 fb$^{-1}$ (from Ref. \protect\cite{0-48}).}
\end{table}

\begin{table}[htb]
\centering
\begin{tabular} {|c|c|c|c|} \hline
$ M_{h_1}$ (GeV) & 80 & 100 & 120   \\ \hline
BR~$(h_1\to \gamma\gamma$) & $0.22\%$ & $0.37\%$ & $0.40\%$  \\ \hline
$S/\sqrt B$    &6.1  & 10.9 & 11.5\\ \hline
 \end{tabular}
\caption{Signals for process $pp\to h_1X$, $h_1\to \gamma\gamma$,
with maximum enhancement, for ${\cal L}$=30 fb$^{-1}$.}
\end{table}    

In Model I$'$ it is not possible
to simultaneously suppress the $h_1b\overline b$ coupling and keep the
 $h_1t\overline t$ coupling
near $\phi^0$ strength. A consequence of this is that although similar
 enhancements of BR~$(h_1\to \gamma\gamma)$ are again 
possible, the production mechanism $(gg\to h_1)$ would be heavily
 suppressed.
However, a chance of detection remains in the associated production channel
$q\overline q\to Wh_1$ with subsequent decays $W\to l\nu_l$ and 
$h_1\to \gamma\gamma$.
Enhancing $h_1\to \gamma\gamma$ would require $\alpha\to \pi/2$,
 $\beta\approx \pi/4$ (i.e. $\beta$ at its smallest value), and
hence $\sin^2(\beta-\alpha)\approx 0.5$; thus the associated production 
cross--section would be suppressed relative to $\phi^0$ by a factor 0.5. 
However, with the
higher BR($h_1\to \gamma\gamma$) one would expect a comparable number of events,
and since
a reasonable statistical signal in the associated production channel is 
possible for
$\phi^0$ with high luminosity, then an analogous signal for $h_1$ may be
obtainable. 
Table~4 uses
the results of the simulation done in Ref. \cite{EAGLE}.  
\begin{table}[htb]
\centering
\begin{tabular} {|c|c|c|c|} \hline
$ M_{\phi^0}$ (GeV) & Signal (S) & Background (B)& $S/\sqrt B$  \\ \hline
80 & 13.5  & 10.5 &  4.2   \\ \hline
110 & 18.3  & 7.0 & 6.9 \\ \hline
\end{tabular}
\caption{Signals for process $pp\to \phi^0lX$, $\phi^0\to
 \gamma\gamma$, for ${\cal L}=100$ fb$^{-1}$ (from 
Ref. {\protect \cite{EAGLE}}).}
\end{table}     

\begin{figure}[htb]
\centerline{\protect\hbox{\psfig{file=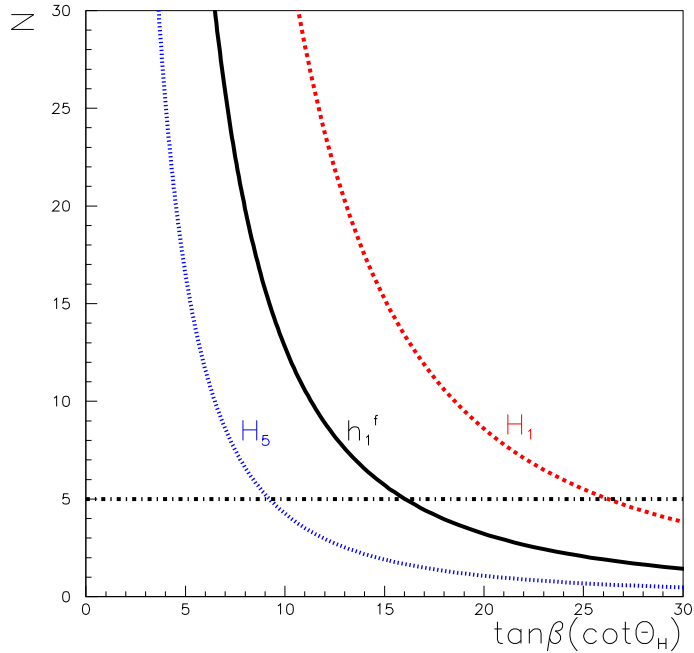,height=10cm,width=10cm}}}
\caption{Statistical signal as a function of $\tan\beta$ for the three
$H_F$ considered, with $M_F=80$ GeV.}
\label{Ferm}
\end{figure}
\begin{figure}[hbt]
\centerline{\protect\hbox{\psfig{file=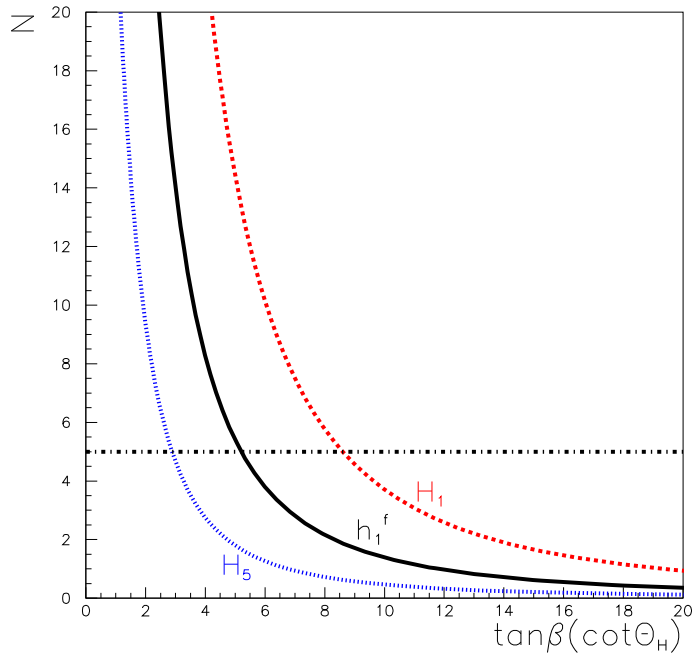,height=10cm,width=10cm}}}
\caption{Same as for Fig.~2 but with $M_F=110$ GeV.}
\label{Ferm3}
\end{figure}
The event numbers for $\phi^0$ include contributions from the 
associated process $pp\to t\overline t\phi^0$, $\phi^0\to \gamma\gamma$,
 with the
lepton trigger originating from $t\to Wb\to l\nu_lb$. 
This contributes 
about $60\%$ of the signal in Table~4. For the above mentioned case
of $h_1$, the contribution in this channel would be negligible (since the
parameter choice for a larger BR~($h_1\to \gamma\gamma$) requires
suppression of the $h_1b\overline b$ and $h_1t\overline t$ coupling).
Therefore we would expect around half the number of events shown in Table~4,
giving a small signal between $2.5\sigma$ and $3.5\sigma$. 
Hence a null search 
in the $gg\to h_1\to \gamma\gamma$ channel but a signal in
the associated channel would then be evidence for Model I$'$.

We recall that a fermiophobic Higgs ($H_F$) may have a sizeable 
BR~$(H_F\to \gamma
\gamma$) \cite{Diaz}. It too has no production channel $gg\to H_F$, but 
would give a clear, distinct signal in the associated
production channel for $H_F\to \gamma\gamma$ decays. 
It has a larger BR~$(H_F\to \gamma\gamma)$ than is possible
in Model I$'$, varying from $20\%$  for $M_{F}=100$ GeV to $\approx 1\%$ 
 for  $M_{F}=130$ GeV. 
 The current bound on $M_F$ from the Tevatron ($\ge 81$ GeV) assumes
 $\phi^0$ strength couplings to vector bosons, which in general is not 
 the case. Hence an $H_F$ with a mass considerably less than 81 GeV is
 not ruled out.
 Figs.~2 and 3 show the statistical significance (defined
by $N=S/\sqrt B$) for two different values of $M_F$ as a function of 
$\tan\beta$($\cot\theta_H$) for the
three $H_F$ considered in Ref. \cite{Ake4}. 
For the Higgs triplet model one must
make the replacement $\cot\theta_H\to\tan\beta$ in the figures. 
The horizontal line at $N=5$ marks the $5\sigma$ discovery limit.
The $H_F$
of the 2HDM (Model I), which requires $\cos\alpha\to 0$, is labelled as 
$h_1^F$ and its cross-section is proportional to $\cos^2\beta$. 
For $h_1^F$ one registers a $5\sigma$ signal if $\tan\beta\le 16$ (5) for
$M_F=80$ GeV (110). For values of $\tan\beta$ close to 1 the signal is of
order $500\sigma$ in Fig.~2 and $70\sigma$ in Fig.~3.     
For $H_1^0$$'$ the coverage is better due to its cross-section being enhanced
by a factor 8/3 relative to that of $h_1^F$. 
When BR~($H_F\to \gamma\gamma)\le 1\%$ (for $M_F\ge 130$ GeV) there will
little difference between the signal for $\phi^0$ and the signal for 
$H_F$. To our knowledge there has not been a rigorous simulation for
the associated production process for heavier Higgs masses, but we may
conclude that detection of $H_F$ will be very difficult in this channel for
$M_F\ge 130$ GeV. 

\subsection{$h_1\to \tau^+\tau^-$}
The next channel we wish to consider is $h_1\to \tau^+\tau^-$. For the
SM Higgs in the
mass region 100 GeV $\le M_{\phi^0}\le 170$ GeV this decay varies from BR=$8\%$ 
to $0.1\%$. One can see from Table~1 that choosing
large $\tan\beta$ (thus $1/\cos\beta$ large) enhances the $h_1ll$ vertex.
\begin{figure}[htb]
\centerline{\protect\hbox{\psfig{file=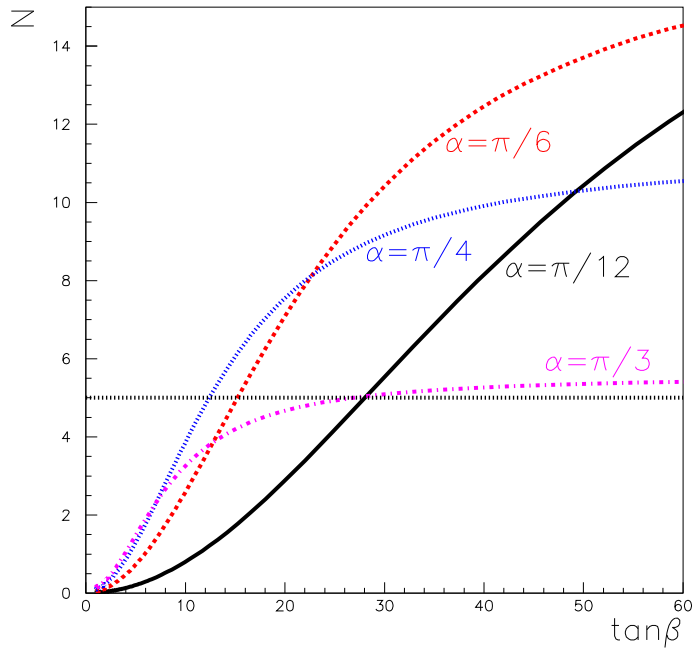,height=10cm,width=10cm}}}
\caption{Statistical signal in the $\tau\tau$ channel as a function of 
$\tan\beta$ for $M_{h_1}=120$ GeV and four different values of $\alpha$.}
\label{Ferm}
\end{figure}
\begin{figure}[htb]
\centerline{\protect\hbox{\psfig{file=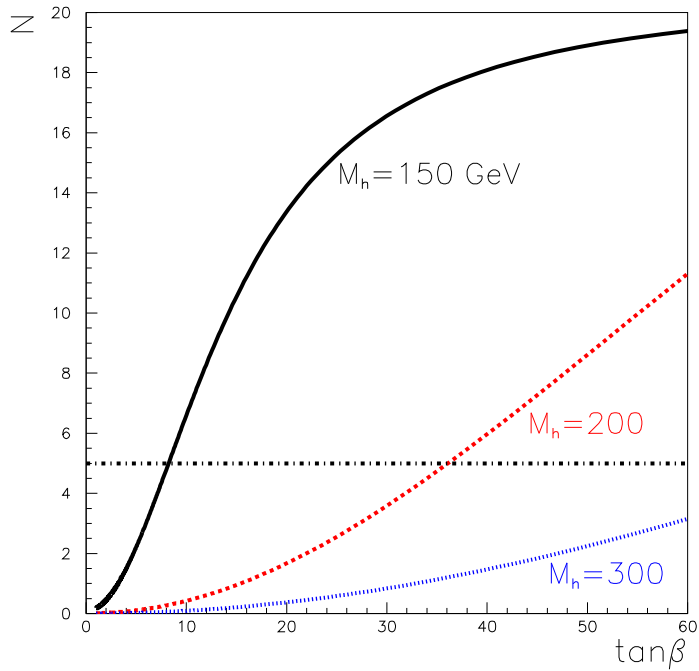,height=10cm,width=10cm}}}
\caption{Same as for Fig.~4 but fixing $\alpha=\pi/4$ and for three different
values of $M_{h_1}$.}
\label{Ferm}
\end{figure}

In Fig.~4 we plot the statistical signal as a function of $\tan\beta$ for
$M_{h_1}=120$ GeV, for four different values of $\alpha$. The choice of
$\alpha=\pi/4$ allows the best coverage, giving a $5\sigma$ signal for
$\tan\beta\ge 12$.
As one moves away from this value the signal can still be strong, although
larger values of $\tan\beta$ are required. As $\tan\beta$ becomes
very large one can obtain BR$(h_1\to\tau\tau)\to 100\%$ for all the
displayed values of
$\alpha$, and thus all four curves will asymptote to a fixed value of $N$. 
Smaller $\alpha$ is favourable from the point of view of the cross--section
and this is why the curves with $\alpha\le \pi/4$ asymptote to larger values
of $N$ than is possible for $\alpha=\pi/4$; they take longer to reach
the $5\sigma$ signal because the BR to $\tau\tau$ for smaller $\tan\beta$
is inferior to the corresponding BR with $\alpha=\pi/4$.
Since setting $\alpha=\pi/4$ gives the best coverage we use this value in
Fig.~5, which shows the expected signal for $M_{h_1}$ 
up to 300 GeV. It is clear from the graphs
that a good signal in this channel is possible even for relatively large
values of $M_{h_1}$.

We note that the enhancement of the Higgs bremsstralung off
a $b$ quark at high $\tan\beta$ is possible in Model II, while 
 BR~$(h_1\to \tau^+\tau^-)$ would approach $10\%$ in this limit.
The combination of an enhanced production cross--section and a 
 BR($h_1\to\tau^+\tau^-$) of
 order $10\%$ would also allow a large signal in the $\tau\tau$ channel.
With values
of $\tan\beta\ge 50$ one could obtain a signal
larger than is obtained for $h_1$ of Model I$'$ using the same $\tan\beta$
 value. However, the accompanying decays in Model II
would be to $b\overline b$ with the process $h_1\to ZZ\to llll$
 suppressed, and so one would not see a signal in the $llll$ channel.
 For Model I$'$ the accompanying decays are $h_1\to WW^*$, $ZZ^*$ and so
 an observable signal in the $llll$ channel is still likely, unless 
 BR~($h_1\to \tau^+\tau^-$)
 is close to $100\%$. Therefore a positive signal in both channels would be
 distinctive of Model I$'$. 

\subsection{$h_1\to t\overline t$, $h_1\to b\overline b$}
Decays of a Higgs to quark pairs, although usually
dominant for the intermediate mass Higgs boson (i.e. $M_{\phi^0}
\le 2M_W$), are 
considered difficult at a hadron collider
and other channels give better chances of detection. However,
 in the 2HDM it is possible that
the quark decays ($t\overline t$ or $b\overline b$) dominate
the vector boson channels over a wide range of Higgs masses and therefore
it is important to infer whether these quark decays can indeed present a 
signature, or if $h_1$ would be hidden.

In any 2HDM it is possible for $h_1\to t\overline t$ decays to predominate
(for $M_{h_1}\ge 2m_t$) if the $VV$ decays are sufficiently
suppressed i.e. $\sin^2(\beta-\alpha)\to 0$. In the extreme case of 
 $\sin^2(\beta-\alpha)=0$ (i.e. $\beta=\alpha$), $t\overline t$ decays would predominate for
 $\tan\beta\le 5$ in all models, while for $\tan\beta\ge 5$ decays to $b\overline b$ would
 predominate in Models II and II$'$.

 We present the event numbers for $h_1$ in Table~5 by 
rescaling the ATLAS numbers, taking $\alpha=\beta$ and 
BR~$(h_1\to t\overline t)=100\%$.
\begin{table}[htb]
\centering
\begin{tabular} {|c|c|c|c|} \hline
$M_{h_1}$ (GeV) & Signal (S) & Background (B) & $S/\sqrt B$ \\ \hline
370 & 1800 & 68600 & 6.9\\ \hline
400 & 1980 & 85700 & 6.8 \\ \hline
500 & 1670 & 127400 & 4.7   \\ \hline
\end{tabular}
\caption{Statistical signal in the $h_1\to t\overline t$ channel for 
${\cal L}=30$ fb$^{-1}$ (from Ref. {\protect \cite{0-74}} with rescaling).}
\end{table}

The statistical significance of the signal is large but
is only meaningful if the theoretical error on $\sigma (pp\to 
t\overline t$) is less than $1\%$, which is not the case at the present.
If the error is reduced then there would be some chance of detection in this
channel. 

The other quark decay that could dominate is that of $h_1\to b\overline b$
in Models II and II$'$ for $\tan\beta\ge 5$. For detection in
this decay mode one requires a good trigger in order to
suppress the huge QCD background.  However, producing this 
$h_1$ in association with a $W$ boson via
$q\overline q\to W^*\to Wh_1$, with $W\to l\nu_l$ (a lepton trigger)
and $h_1\to b\overline b$ only
probes $M_{h_1}\le 120$ GeV  \cite{0-74}, \cite{0-43}. 
Therefore we conclude that the quark decays ($t\overline t$, $b\overline b$)
of $h_1$ would be difficult to observe at the LHC.

\subsection{$h_1\to ZZ^{(*)}$}
The SM Higgs decay to two $Z$ bosons, with the subsequent
decay $ZZ\to llll$ is the `gold--plated' 
channel, since it gives
a large signal throughout the range 130 GeV$\le M_H\le 800$ GeV. Below the
real $ZZ$ threshold one of the vector bosons must be off--shell. In the
2HDM, $\Gamma(h_1\to VV^{(*)}$) is suppressed by $\sin^2(\beta-\alpha)$ 
although for $M_{h_1}\ge 150$ GeV
these decays will usually predominate, unless this suppression is large and/or
another channel is greatly boosted. Table~6 shows the expected signal
 and background ratio for $\phi^0$, which is very large over much of the
mass interval 200 GeV$\le
M_{\phi^0}\le 800$ GeV, reaching as high as 40 standard deviations.
\begin{table}[htb]
\centering
\begin{tabular} {|c|c|c|c|} \hline
$M_{h_1}$ (GeV) & Signal (S) & Background (B) & $S/\sqrt B$ \\ \hline
120 & 5.2 & 4.7 & 2.4 \\ \hline
130 & 24.8 & 8.2 & 8.5 \\ \hline
150 & 68.5 & 10.0 & 21.7 \\ \hline
170 & 19.9 & 9.5 & 6.5 \\ \hline
180 & 51.9 & 9.0 & 17.3 \\ \hline
200  & 189 & 29& 35.3 \\ \hline
300 & 314 & 68 & 38.2 \\ \hline
400& 267 & 56 & 35.7   \\ \hline
500 & 137 & 29 & 25.6 \\ \hline
600 & 70 & 25 & 14.1 \\ \hline
700 & 38 & 21 & 8.3  \\ \hline
800 & 22 & 17 & 5.4 \\ \hline
\end{tabular}
\caption{Statistical signal in the $\phi^0\to ZZ^{(*)}\to llll$ channel for 
${\cal L}=100$ fb$^{-1}$ (from  Ref.{\protect \cite{0-48}}.}
\end{table}
   
For the mass region 100 GeV$\le M_{\phi^0}\le 130$ GeV the $ZZ\to llll$ channel
does not provide a strong enough signature since 
the BR($\phi^0\to llll$) is too low. In a
 2HDM one can suppress
  the $b\overline b$ channel in this region
which may enable the $llll$ signal to be used for lower masses of $h_1$
than is possible for $\phi^0$.
 An extreme case of this is the fermiophobic Higgs ($H_F$) which has
a considerably larger BR to $ZZ^*$ than is possible for $\phi^0$ 
in this mass region.
 Enhancements are possible in the other versions of the 2HDM, although
one cannot suppress
all fermion channels simultaneously -- an option only available in Model I.
A major problem with detecting $H_F$ is that there is no $gg\to H_F$ 
production process,
and $VV$ fusion will be suppressed by at least a factor of 0.5
 (for $\alpha=\pi/2$, $\beta=\pi/4$).
This results in a sizeable drop in the cross-section, $\sigma(pp\to
 H_FX$), by at least a factor of 10 relative to $\phi^0$ 
in the range 80 GeV$\le M_{H_F}\le 120$ GeV.  However,
BR~($H_F\to ZZ^*$) is much larger than that for $\phi^0$,
 the former varying from $6\%\to 9\%$ ($20\%\to 29\%$ 
 for $H^0_5$) in the region 80 GeV$\le M_{H_F}\le 120$
 GeV. This is in contrast to
BR~$(\phi^0\to ZZ^*$) which varies from $0.01\% \to 1.3\%$.
Therefore the signal event number may be considerably greater
(an order of magnitude) for $H_F$ in the region 
$M_F\le 100$ GeV than for $\phi^0$. In fact, a signal for $\phi^0$ in this channel 
is unlikely for $M_{\phi^0}\le 120$ GeV (see Table~6), but may be possible
for $H_F$ with a few events over a small background.
For $M_F\ge 120$ GeV the signal for $H_F$ becomes weaker than that for
$\phi^0$ since BR~$(\phi^0\to ZZ^*)$ is increasing rapidly, and coupled with 
the latter's superior cross--section more events are produced overall.  
 
Other models which may provide an enhanced signal in the $llll$
channel are Models II and II$'$, although Model II has the possibility
of the largest signal since both 
$b\overline b$ and $\tau^+\tau^-$ can
be simultaneously suppressed. The choice of $\cos^2\alpha\to 1$
and moderate $\sin^2\beta$ ($\approx 0.5$) would boost the $t\overline t$
coupling and so enhance the $gg\to h_1$ cross--section, and also cause the
required
suppression of the $b\overline b$ and $\tau^+\tau^-$ decays (in Model II$'$
the $\tau^+\tau^-$ cannot be simultaneously suppressed, and so
the BR to $ZZ^*$ would be less). This means
that although $c\overline c$ and $gg$ decays would dominate,
 BR~$(h_1\to ZZ^*)=1\%$ is possible, and combined with the enhanced
  cross--section one 
would find an increase of order 4 for the $llll$ signal. This analysis is
for $M_{h_1}=110$ GeV. 

For the heavier Higgs masses, $M_{h_1}\ge 2M_Z$, the decay to two real
$Z$ bosons $h_1\to ZZ$ is available.  Two signatures remain which 
could suggest a 2HDM in this channel:
 
\begin{itemize}

\item[{(i)}] An enhanced cross--section. Since $gg\to \phi^0$
gives the largest rate, if one enhances the $h_1t\overline t$ coupling it
is possible to have more $llll$ events than for the $\phi^0$ case.
  
\item[{(ii)}] A suppression of the $llll$ signal relative to $\phi^0$
would be very noticeable, since the $\phi^0$ signal can be as large 
as $40\sigma$.

\end{itemize}
Case (i) is possible in all 2HDM, as mentioned earlier for parameter
choices such as $\cos^2\alpha\to 1$ and moderate $\sin^2\beta$ ($\approx 0.5$).
This would cause an enhancement of 2 for the $gg$ fusion cross--section, and
the overall cross--section would be enhanced by roughly the same factor
(since $gg$ fusion gives the largest contribution).
Case (ii) is also possible in any 2HDM and therefore the
above scenarios do not give any information
on which particular model would be present. There are two ways for case (ii) to be 
realized.
\begin{itemize}

\item[{(a)}] The choice of $\beta\approx\alpha$ forces $\sin^2(\beta-\alpha)
\approx 0$ and so the $h_1\to ZZ$, $WW$ decays are heavily suppressed.
 Instead the $t\overline t$ or $b\overline b$ channel predominates for 
 the heavy Higgs mass region, $M_{h_1}\ge 2M_Z$ (see Section 3.3).  
  
\item[{(b)}] The BRs to $ZZ$ and $WW$ may be kept dominant, but with
 the production cross--section heavily suppressed, i.e. the $h_1t\overline t$
coupling is heavily reduced which in turn suppresses the mechanism $gg\to h_1$.
\end{itemize}  

Case (b) is true for a fermiophobic Higgs whose main production
process would be 
$WW$ and $ZZ$ fusion. Using the ATLAS studies and Table~6 
we can evaluate the statistical signal. The cross--section 
is roughly 10 times less for $h^F_1$ than for $\phi^0$ throughout the
range 200 GeV$\le M_{H_F}\le 800$ GeV, and BR~$(h^F_1\to ZZ)\approx$ 
BR~$(\phi^0\to ZZ)$. We would find that 
a 4$\sigma$ signal is only possible up to 
$M_{F}=400$(600) GeV for $h_1^F(H^0_1$$'$). For $H^0_5$ the cross--section
is roughly half that of $h_1^F$ throughout the heavy mass region,
 but this is compensated by the fact that BR($H^0_5\to ZZ$) is twice
 that of $h_1^F$, with the result being that a similar signal is expected. 
 This analysis is for maximum values for production cross--sections.
 Therefore it turns out that there is a sizeable parameter space 
 ($\cos^2\beta$, $M_{F}$) for a hidden $H_F$ at the LHC.
  We stress that the distinctive signature
of a fermiophobic Higgs ($H_F\to \gamma\gamma$) is lost in this heavier mass
region, and the suppressed $llll$ signal could be mimicked by another
$h_1$ with suppressed $ZZ$ decays and/or production cross--section. 

\section{Conclusions}
We have studied the phenomenology of the lightest CP--even neutral Higgs
boson($h_1$) of a non--SUSY, non--minimal Standard Model at the LHC. 
Emphasis was given to the problem of distinguishing $h_1$ from the
SM Higgs boson ($\phi^0$) and to the detection prospects of a 
fermiophobic Higgs ($H_F$). This
is complementary to our earlier work which considered prospects at LEP2.
 We considered
four different decay channels and showed that the following signals
are possible:

\begin{itemize}

\item[{(i)}] An enhanced signal in the $pp \to h_1X$, $h_1\to
\gamma\gamma$ 
channel would
be evidence for Model II or II$'$, although Model II can produce the greater
number of events. A signal in the associated production channel 
$pp \to h_1W$, $h_1\to
\gamma\gamma$ and $W\to l\nu_l$ but no signal in the above channel would be
evidence for Model I$'$ or a fermiophobic Higgs, with the latter capable
of more signal events.  
\item[{(ii)}]
A large signal from $h_1\to \tau^+\tau^-$ decays is possible in Model I$'$,
 even for relatively large masses, and would be accompanied by a 
  signature in the $llll$ channel, unless the former has
 a branching ratio close to $100\%$. In Model II it is
 possible to have a comparable enhancement of $\tau^+\tau^-$ events but the
 accompanying decays would be to hadrons, the latter unlikely to
 be separated from the backgrounds.  
\item[{(iii)}] Enhanced $h_1\to t\overline t$, $b\overline b$ decays in the 
heavy Higgs
region are possible although they are very difficult to detect at the LHC 
due to the
 large jet background. Detection may be possible in the 
 $t\overline t$ channel
 if the theoretical error on the background $pp\to t\overline t$
  cross--section can be reduced to below $1\%$.
\item[{(iv)}] A suppressed or enhanced signal in the $llll$ channel
 would indicate a non--minimal Higgs sector, although would not
shed light on which model were actually present.
  
Detection of a fermiophobic Higgs will be difficult for masses greater
than 130 GeV, the only chance being a suppressed signal in the $llll$ channel.
Such a suppression, however, may be mimicked in all the models considered
and so is not indicative of fermiophobia. The unmistakable signature of $H_F$  
is available up to $M_F\le 130$ GeV, and would be a spectacular
excess of $\gamma\gamma$ events in the associated production channel
provided that the cross-section is not so suppressed.

\end{itemize}

\section*{Acknowledgements} 
This work has been supported by the UK Royal Society.

\end{document}